\documentclass[12pt,preprint]{aastex}
\usepackage{aastexug}
\usepackage{graphicx}

\begin{document}

\title{An Analytic Model of Black Hole Evolution and Gamma Ray Bursts}

\author{Ding-Xiong Wang\altaffilmark{1,2}, Wei-Hua Lei\altaffilmark{1}, Kan Xiao\altaffilmark{1}, Ren-Yi Ma\altaffilmark{1}}
\affil{Department of Physics, Huazhong University of Science and Technology, Wuhan, 430074,China}
\email{dxwang@hust.edu.cn}

\altaffiltext{1}{Department of Physics, Huazhong University of Science and Technology, Wuhan, 430074, China}
\altaffiltext{2}{Send of\/fprint requests to: Ding-Xiong Wang (dxwang@hust.edu.cn)}

\label{firstpage}

\begin{abstract}
An analytic model of the evolution of a rotating black hole (BH)
is proposed by considering the coexistence of disk accretion with
the Blandford-Znajek process.  The evolutionary characteristics of
the BH are described in terms of three parameters: the BH spin
$a_*$, the ratio $k$ of the angular velocity of the magnetic
f\/ield lines to that of the BH horizon and the parameter
$\lambda$ indicating the position of the inner edge of the disk.
It is shown that the ratio $k$ being a little greater than 0.5
af\/fects the evolutionary characteristics of the BH
signif\/icantly, and the BH spin increases rather than decreases
in its evolutionary process provided that the initial value of the
BH spin is located in an appropriate value range determined by
ratio $k$. Our calculations show that the system of a BH accretion
disk with $k=0.6$ might provide a much higher output energy in a
shorter timescale for gamma-ray bursts than the same system with
$k=0.5$.
\end{abstract}


\section{INTRODUCTION}
As is well known, the Blandford-Znajek(BZ) process was proposed
originally as a possible energy mechanism of quasars and active
galactic nuclei(AGNs;Blandford \& Znajek 1977; Macdonald \& Thorne
1982, hereafter MT82; Rees 1984). Recently, Lee, Wijers, \& Brown
(2000) proposed that the BZ process can be used as a central
engine for powering gamma-ray bursts (GRBs), where the rotating
energy of a stellar black hole (BH) with a magnetic f\/ield of
$10^{15} G$ is extracted along the magnetic f\/ield lines
supported by a transient magnetized accretion disk. Very
 recently,
Lee and Kim (2000, 2002) proposed a model of the evolution of a
rotating BH at the center of GRBs by considering the ef\/fects of
the BZ process with the transient disk (hereafter the LK model),
and some constraints to the parameters of the BH accretion disk
are given.

In this paper an analytic model of BH evolution is proposed based
on the LK model and is improved in two respects: (1) A parameter
$\lambda$ is introduced for the position of the inner edge of the
disk/torus, which is located between the innermost bound orbit and
the last stable orbit. (2) Another parameter $k$ is used to
indicate the ratio of the angular velocity of the magnetic f\/ield
lines to that of the BH horizon. It is shown that the ratio $k$
being a little greater than 0.5 af\/fects the evolutionary
characteristics of the BH signif\/icantly, and the BH spin
increases rather than decreases in its evolutionary process
provided that the initial value of the BH spin is located in an
appropriate value range determined by the ratio $k$. Our
calculations show that the system of a BH accretion disk with
$k=0.6$ might provide a much higher output energy in a shorter
timescale for GRBs than the same system with $k=0.5$.

This paper is organized as follows: In \S2 we derive the basic
equations of BH evolution and the corresponding characteristic
functions by introducing two parameters $\lambda$ and $k$. In
\S\S3 and 4 we discuss the ef\/fects of the parameters $k$ and
$\lambda$ on BH evolution and GRBs by using the parameter space
and the corresponding characteristic functions, respectively.
Finally, in \S5 we discuss some problems concerning our model.
Geometrized units ($G=c=1$) are used in this paper.

\section{BASIC EQUATIONS OF BH EVOLUTION AND CHARACTERISTIC FUNCTIONS}
Now we are going to describe our model for the evolution of a
rotating BH surrounded by a magnetized accretion disk based on the
LK model. First, a parameter $\lambda$ is introduced to indicate
the position of the inner edge of the disk in the following
equation:
\begin{equation}
\chi_{in}=\chi_{mb}+\lambda (\chi_{ms}-\chi_{mb}), \ \ \ 0 \le \lambda \le 1,
\end{equation}
where $\chi_{in}$ is the dimensionless radial parameter
corresponding to the inner-edge radius $r_{in}$ of the disk, which
is located between $r_{mb}$ and $r_{ms}$ (Abramowicz, Jaroszynski,
\& Sikora 1978; Abramowicz \& Lasota 1980). The quantities
$\chi_{in}, \chi_{mb}$, and $\chi_{ms}$ are dimensionless radial
parameters:
\begin{equation}
\chi_{in} \equiv \sqrt{\frac{r_{in}}{M}}, \ \ \ \chi_{mb} \equiv
\sqrt{\frac{r_{mb}}{M}}, \ \ \ \chi_{ms} \equiv
\sqrt{\frac{r_{ms}}{M}}.
\end{equation}
 It is found from equation (1) that the inner edge of a thick disk varies continuously from the innermost bound
orbit to the last stable orbit as $\lambda$ changes from zero to unity, and the disk becomes a thin disk as $\lambda$
attains unity.

Usually the evolutionary state of a Kerr BH can be described by
two parameters: BH mass $M$ and spin $a_*$, and the latter is
related to the BH mass $M$ and angular momentum $J$ by
\begin{equation}
a_* \equiv \frac{J}{M^2}, \ \ \ 0 \le a_* <1.
\end{equation}
Both $\chi_{mb}$ and $\chi_{ms}$ depend only on $a_*$  (Novikov \&
Thorne 1973; Abramowicz et al. 1978):
\begin{equation}
\chi_{mb} = 1+\sqrt{1-a_*},
\end{equation}
\begin{equation}
\chi_{ms}^4-6\chi_{ms}^2+8 a_* \chi_{ms} - 3 a_*^2=0.
\end{equation}
The specif\/ic energy $E_{in}$ and the specif\/ic angular momentum
$L_{in}$ corresponding to the inner-edge radius $r_{in}$ of the
disk/torus are expressed by (Novikov \& Thorne 1973)
\begin{equation}
E_{in}=\frac{(1-2\chi_{in}^{-2}+a_* \chi_{in}^{-3})}{(1-3 \chi_{in}^{-2}+2a_*\chi_{in}^{-3})^{1/2}},
\end{equation}
\begin{equation}
L_{in}=\frac{M \chi_{in} (1-2a_* \chi_{in}^{-3}+a_*^2
\chi_{in}^{-4})}{(1-3\chi_{in}^{-2}+2a_*\chi_{in}^{-3})^{1/2}}.
\end{equation}
In the LK model, the ef\/fects of the BZ process on the evolution
of a rotating BH are taken into account in the environment of a
transient magnetized accretion disk. Due to the lack of knowledge
of the remote astrophysical load in the BZ process, it is doubtful
that the matching condition $k=0.5$ can be satisf\/ied exactly.
Therefore,  we introduce a parameter $k$ to describe the rotation
of the magnetic f\/ield lines relative to the spinning BH,
\begin{equation}
k \equiv \frac{\Omega_f}{\Omega_h}, \ \ \ 0<k<1,
\end{equation}
where $\Omega_f$ is the angular velocity of the magnetic f\/ield
lines and $\Omega_h$ is that of the horizon and is related to the
horizon radius $r_{_h}$ by
\begin{equation}
\Omega_h=\frac{a_*}{2 r_{_h}},\ \ \ r_{_h}=M(1+q).
\end{equation}
By using a modif\/ied equivalent circuit for the BZ process based
on MT82, we derive the expression for the rate of extracting
energy from the rotating BH in the BZ process (hereafter the BZ
power; Wang, Xiao, \& Lei 2002) as follows:
\begin{equation}
P_{BZ}=4k(1-k)P_{BZ}^{optimal}, \end{equation}
\begin{equation}
P_{BZ}^{optimal}=B_h^2M^2[Q^{-1}arctan Q -\frac{1+q}{2}],
\end{equation}
\begin{equation}
Q \equiv \sqrt{\frac{1-q}{1+q}},\ \ \ q=\sqrt{1-a_*^2},
\end{equation}
where $P_{BZ}^{optimal}$ is the optimal BZ power corresponding to
the impedance matching with $k=0.5$ and $B_h$ is the magnetic
f\/ield on the horizon. Both $Q$ and $q$ are the parameters
depending only on the BH spin. Taking $k=0.5$ in equation (10), we
have the same expression for the optimal BZ power as given by Lee
et al. (2000), who pointed out that the BZ power had been
underestimated by a factor of 10 in previous works.

Based on the conservation of energy and angular momentum, the
basic evolutionary equations of a rotating BH with disk accretion
by the BZ process can be written as (Park  \& Vishniac 1988;
Moderski \& Sikora 1996; Moderski, Sikora, \& Lasota 1997; Wang,
Lu, \& Yang 1998; Lee \& Kim 2000)
\begin{equation}
\frac{dM}{dt}=E_{in}{\dot M}_d-P_{BZ},
\end{equation}
\begin{equation}
\frac{dJ}{dt}=L_{in}{\dot M}_d-\frac{P_{BZ}}{k\Omega_h}.
\end{equation}
Incorporating equations (13) and (14), we obtain the evolutionary
equation of the BH spin:
\begin{equation}
\frac{da_*}{dt}=M^{-2}(L_{in}{\dot M}_d-\frac{P_{BZ}}{k
\Omega_h})-2M^{-1}a_*(E_{in}{\dot M}_d-P_{BZ}).
\end{equation}

The accretion rate ${\dot M}_d$ of the transient magnetized disk
is related to $B_h$ by (Lee \& Kim 2000)
\begin{equation}
{\dot M}_d=\frac{M r_{in}^3}{\varpi^2
r_h^2}M^2a_*^2B_h^2=\frac{B_h^2M^2Q^2\chi_{in}^2}{1+(a_*/\chi_{in}^2)^2(1+2/\chi_{in}^2)},
\end{equation}
where $\varpi$ is the Kerr metric parameter  at the inner edge.
According to the LK model, $B_h$ depends on the mass loss from the
disk  (Lee \& Kim 2000):
\begin{equation}
B_h^2=B_h^2(0)D(t),
\end{equation}
\begin{equation}
D(t)=1-\frac{\int_{0}^{t}{\dot M}_ddt}{M_d(0)},
\end{equation}
where $M_d(0)$ and $B_h(0)$ are the initial disk mass and magnetic
f\/ield, respectively.

Substituting the above expression for $E_{in}$, $L_{in}$,
$\chi_{in}$, $\Omega_h$, $P_{BZ}$ and ${\dot M}_d$ into equations
(13) and (15), we obtain the following evolutionary equations:
\begin{equation}
\frac{dM}{dt}=f(a_*,k,\lambda){\dot M}_d,
\end{equation}
\begin{equation}
\frac{da_*}{dt}=M^{-1}g(a_*,k,\lambda){\dot M}_d.
\end{equation}
From equations (19) and (20), we f\/ind that $dM/dt$ and $da_*/dt$
have the same sign as $f(a_*,k,\lambda)$ and $g(a_*,k,\lambda)$,
and hereafter the two functions are referred to as the
characteristic function of BH mass (CFBHM) and that of BH spin
(CFBHS), respectively.

\section{EFFECTS OF THE PARAMETER $k$ ON BH EVOLUTION AND GRBs}
The characteristics of BH evolution can be well described by using
the CFBHM and CFBHS, and the curves of these two functions varying
with $a_*$ for a thin disk ($\lambda=1$) with dif\/ferent values
of $k$ are shown in Figure 1.
It is shown in Figure 1a that $f(a_*,k,1)$ is negative when
$a_*>0.9981, 0.9984,$ and $0.9999$ for $k=0.5$, 0.6, and 0.8,
respectively. From Figure 1b we f\/ind that $g(a_*,k,1)$ is always
negative for $k=0.5$, while it is positive for $k=0.6$ with
$0.3114<a_*<0.8067$, and for $k=0.8$ with $0.1205<a_*<0.9598$.
These results imply that the BH mass might decrease as the BH spin
approaches unity very closely, and the BH could be spun up in the
duration of powering the GRB, provided that the parameter $k$ is a
little greater than 0.5.

The ef\/fect of the variation of $k$ on the evolution of the BH
can be discussed more visually in the parameter space consisting
of $a_*$ and $k$ with the curves represented by $f(a_*,k,1)=0$ and
$g(a_*,k,1)=0$, as shown in Figure 2.

It is found that the parameter space is divided into three regions
by two thick solid curves, which are represented by $g(a_*,k,1)=0$
and $f(a_*,k,1)=0$. The former is the common boundary of regions I
and II (curve \textbf{\textsl{abc}}), as shown in Figure 2a, and
the latter is the common boundary of regions I and III, as shown
in Figure 2b. Regions I and III are each further divided into two
sub-regions, IA and IB, and IIIA and IIIB, by the horizontal thin
line \textbf{\textsl{bd}}, as shown in Figures 2a and b,
respectively. The segment \textbf{\textsl{bd}} is tangent to the
curve $g(a_*,k,1)=0$ at the bottom point
\textbf{\textsl{b}}(0.5610, 0.5118). In the parameter space, each
filled circle with an arrow is referred to as a representative
point (RP), which represents one evolutionary state of BH. We can
use RP displacement in the parameter space to describe the
evolution of the BH. According to the positions of RPs, we have
f\/ive dif\/ferent BH evolutionary states corresponding to f\/ive
subregions, as shown in Table 1.

From Table 1 we f\/ind that the f\/ive evolutionary states of a
rotating BH can be easily determined by RP position in the
parameter space, and the details are given as follows:

(1) The RPs in regions IA and IB represent the evolutionary states
of the BH with increasing mass and decreasing spin, and the BH
with the RP in region IA will never stop rotating, while the BH
with the RP in region IB might evolve to a Schwarzschild BH with
zero spin in long enough time. The evolution state given in the LK
model is just described by the RP in region IB.

(2) The RPs in region II represent the evolutionary states of the
BH towards the equilibrium spin with increasing mass and spin
increasing toward equilibrium, which makes the BH a more powerful
central engine for GRBs.

(3) The RPs in regions IIIA and IIIB represent the evolutionary
states of the BH with decreasing mass and decreasing spin, which
implies that the ef\/f\/iciency of transferring  the accreted mass
into GRB energy might be greater than unity in these two
evolutionary states.

(4) As shown in Figure 2 and Table 1, there are two possible
evolutionary terminals for the BH, if the evolution time is long
enough. The terminal with equilibrium spin $a_*^{eq}$ corresponds
to the  RPs in regions IA, II, and IIIA, while that with zero spin
corresponds to the RPs in regions IB and IIIB.

From the above discussion, we f\/ind that evolutionary
characteristics of a rotating BH in the BZ process are very
sensitive to the parameter $k$, especially as $k$ varies across
the vertical ordinate of the bottom point $k^{bot} \approx 0.5118$
in Figure 2a.

The value range of $a_*$ in region II is expressed by
$a_*^{turn}<a_*<a_*^{eq}$, where $a_*^{turn}$ and $a_*^{eq}$ are,
respectively, the left and right intersections of the curve
\textbf{\textsl{abc}} with a horizontal line above the segment
\textbf{\textsl{bd}}. Inspecting Figure 2a, we f\/ind that the
more  the horizontal line is above the segment
\textbf{\textsl{bd}} or the greater $k$  is than $k^{bot}$, the
greater  the value range of $a_*$ is in region II. The left part
of curve \textbf{\textsl{abc}} is the set of $a_*^{turn}$, while
the right part of curve \textbf{\textsl{abc}} is the set of
$a_*^{eq}$. Obviously, $a_*^{turn}$ is unstable, while $a_*^{eq}$
is stable for the evolutionary states.

In order to discuss the ef\/fects of the parameter $k$ on the
evolution of a rotating BH and the related characteristics of
GRBs, we compare the results of the BH accretion disk system by
taking $k=0.6$ and $k=0.5$, respectively. Our main results are
shown in Figures 3-8, in which  the solid lines are used for the
results with $k=0.6$, and the dashed lines for those with $k=0.5$.
In our calculations, the initial values of the concerned
quantities are taken as those given in the LK model:
$M_h(0)=7M_\odot$, $M_d(0)=3M_\odot$, and $B_d(0)=10^{15}G$.

(1) In Figure 3 the BH spin $a_*$ increases with the evolution
time if its initial value is taken to be $0.3114<a_*(0)<0.8067$
for $k=0.6$, while it always decreases with the evolution time no
matter what initial value it takes for $k=0.5$. As shown in Figure
1b, 0.3114 and 0.8067 are exactly the values of $a_*^{turn}$ and
$a_*^{eq}$ for $k=0.6$, respectively.

(2) In Figure 4 $a_*^c$ can never be greater than
$a_*^{turn}=0.3114$ no matter how much the initial disk mass is
for $k=0.6$, while it increases monotonically with the initial
disk mass for $k=0.5$. Since $a_*^c$ is def\/ined as the critical
BH spin to ensure that the BH will stop spinning after all the
initial disk mass accretes onto the BH in the LK model, our result
implies that a BH with its initial spin greater than $a_*^{turn}$
can never be spun down to a Schwarzschild BH.

(3) In Figure 5 the output energy $E$ from a system with $k=0.6$
({\it{solid lines}}) is always  greater than that from a system
with $k=0.5$ ({\it{dotted lines}}). In the former case, $E$ will
monotonically increase with the initial disk mass $M_d(0)$ without
an upper limit, provided that the initial BH spin is greater than
$a_*^{turn}$, while in the latter case, it always keeps constant
for $M_d(0)$ greater than one critical value. Another dif\/ference
between these two systems lies in the derivative of $E$ with
respect to $M_d(0)$. As shown in Table 2 the f\/irst derivative
$dE/dM_d(0)$ always increases with $M_d(0)$ (in units of solar
mass) provided that the system has the parameter $k=0.6$ with
$0.3114<a_*(0)<0.8067$, while it always decreases with $M_d(0)$
either for $k=0.6$ with $0<a_*(0)<0.3114$ or for $k=0.5$ with any
value of $a_*(0)$.  These results can be easily explained by the
RPs in the parameter space (Figure 2a) and the relation of the BZ
power to the BH spin.

(4) In Figure 6 the output energy from the system with $k=0.6$ is
always greater than that with $k=0.5$ for any initial values of
the BH spin. Def\/ining the relative dif\/ference of output energy
between the above two systems as $\delta \equiv
(E_{out}^{0.6}-E_{out}^{0.5})/E_{out}^{0.5}$, we have the curve of
$\delta$ versus $a_*(0)$ as shown in Figure 7, and the maximum is
$\delta_{max} \approx 2.44$ at $a_*(0) \approx 0.294$. Thus, the
output energy for GRBs can be remarkably augmented  by a little
increase above $k^{bot}$.

(5) In Figure 8 the evolutionary timescales $t_{evl}$ of GRBs
versus $a_*(0)$ are obtained by incorporating equations (16) $-$
(18) and the initial values of the systems with $k=0.5$ and
$k=0.6$, where the cutof\/f of $t_{evl}$ is taken as $T_{90}$
def\/ined in the LK model. A peak value of $t_{evl}$ is found for
each system, which is located at $a_*(0) \approx 0.298$ for
$k=0.6$ and at $a_*(0) \approx 0.402$ for $k=0.5$, respectively.
It is shown that the  $t_{evl}$ of the system with $k=0.6$ is
always less than that with $k=0.5$ for $a_*(0)>0.366$. These
results might be helpful to explain GRBs with the most energetic
power.

\section{EFFECTS OF THE PARAMETER $\lambda$ ON BH EVOLUTION AND GRBs}
Taking the parameter $\lambda$ into account, we can discuss the
evolutionary characteristics of a BH surrounded by a thick
disk/torus, and the parameter spaces with boundary curves
$g(a_*,k,\lambda)=0$ and $f(a_*,k,\lambda)=0$ corresponding to
dif\/ferent values of $\lambda$ are shown in Figure 9. It is found
that the above f\/ive evolutionary states of a BH surrounded by a
thin disk $(\lambda=1)$ remain valid  in the case of a thick
disk/torus with its inner edge positioned by the parameter
$\lambda$ $(0 \le \lambda <1)$. It is interesting to note that
both the bottom point of the curve $g(a_*,k,\lambda)=0$ and the
leftmost point of the curve $f(a_*,k,\lambda)=0$ vary
nonmonotonically with the parameter $\lambda$, and we have the
curves of $k^{bot}$ and $a_*^{left}$ versus $\lambda$ as shown in
Figures 10 and 11, respectively. The quantity $k^{bot}$ attains
its maximum of 0.598 at $\lambda \approx 0.137$, and $a_*^{left}$
attains its minimum of 0.995 at $\lambda \approx 0.261$. The
output energy $E_{out}$ from the systems and the evolutionary
timescales $t_{evl}$ corresponding to the dif\/ferent values of
$\lambda$ with $k=0.6$ are compared in Figures12 and 13,
respectively. Finally, the relative dif\/ferences of output energy
def\/ined by $\delta \equiv
(E_{out}^{0.5}-E_{out}^{0.5})/E_{out}^{0.5}$ corresponding to the
dif\/ferent values of $\lambda$ are shown in Figure 14. The main
ef\/fects of the parameter $\lambda$ on BH evolution and GRBs are
summarized as follows.

(1) The BH will be spun up in the case of a thick disk $(0 \le
\lambda <1)$ as well as in the case of a thin disk $(\lambda =1)$,
provided that we have $k>k^{bot}$ with
$a_*^{turn}<a_*(0)<a_*^{eq}$, where $a_*^{turn}$ and $a_*^{eq}$
have the same meaning as given in \S3.

(2) The ef\/f\/iciency of transferring the accreted mass into GRB
energy might be greater than unity provided that the RP of BH
evolutionary state is located in the corresponding region III as
shown in Figure 9b.

(3) From Figures 12 and 13 we f\/ind that the output energy from a
system with a thick disk/torus might be greater, and the concerned
evolutionary timescales might be longer, than the corresponding
values of a system with a thin disk if $a_*(0)$ is greater than
some value. Therefore,  our model with dif\/ferent values of
$\lambda$ might adapt to the GRBs with dif\/ferent output energy
and dif\/ferent timescales.

(4) From Figure 14 we f\/ind that the output energy from a system
with $k=0.6$ is always greater than that from a system with
$k=0.5$, no matter what value  $a_*(0)$ has. This is true for a
system with a thick disk/torus as well as for a system with a thin
disk.

In summary, compared to the ef\/fects of $k$ the characteristics
of BH  evolution and GRBs are not remarkably  af\/fected by
variation of the parameter $\lambda$.

\section{DISCUSSION}
In this paper the evolutionary characteristics of a rotating BH
surrounded by a transient magnetized disk are discussed by
considering the coexistence of disk accretion with the BZ process,
and the two parameters $\lambda$ and $k$ are introduced in our
model to modify the LK model. A main consequence of our model is
that the BH will be spun up rather than spun down provided that
$k$ is greater than a critical value $k^{bot}$ with
$a_*^{turn}<a_*(0)<a_*^{eq}$. Our calculations show that more
energy from a BH system with $k$ above $k^{bot}$ might be
extracted in less time compared to a BH system with $k$ below
$k^{bot}$.

However there are still some problems related to our model. First,
the dif\/ferent values of the ratio $k \equiv \Omega_f/\Omega_h$
play a very important role in our model. Unfortunately, the
determination of this ratio has remained one of the main
unresolved problems in the BZ process. It was argued in a
speculative way that the ratio might be regulated to about 0.5, if
the charged particles conspired with the BH to have just the right
amount of inertia for the impedance matching (MT82). However, it
is difficult to understand how the load can conspire with the BH
to have the same resistance and satisfy the matching condition,
since the load is so far from the BH that it cannot be causally
connected (Punsly \& Coroniti 1990).

It was pointed out in the theory of BH magnetospheres that
$\Omega_f$ can be regarded as a function of magnetic f\/lux
$\Psi$, i.e., $\Omega_f(\Psi)$, since it is constant on the
magnetic surface due to isorotation and axial symmetry (MT82).
Very recently, Beskin and Kuznetsova (2000) discussed the stream
equation describing magnetic surfaces $\Psi (r,\theta)$, and found
that $\Omega_f$ can be either greater or less than $\Omega_h/2$
for the dif\/ferent possibilities of the f\/low, with the electric
current f\/ixed by the pair creation region in the BH
magnetosphere. However, we  still have a long way to go for the
determination of $\Omega_f$ in a BH magnetosphere.

Second, we notice that the relation in equation(16) between ${\dot
M}_d$ and $B_h$ is the key to af\/fecting the evolutionary
characteristics. To illustrate this, we replace equation (16) with
another relation between ${\dot M}_d$ and $B_h$, which was
proposed based on the balance between the pressure of the magnetic
f\/ield on the horizon and the ram pressure  of the innermost
parts of the accretion f\/low (Moderskin et al. 1997), i.e.,
\begin{equation}
\frac{B_h^2}{8 \pi}=P_{ram} \sim \rho c^2 \sim \frac{{\dot M}_d}{4
\pi r_h^2}.
\end{equation}
From equation (21) we assume the relation to be
\begin{equation}
{\dot M}_d=\frac{r_h^2B_h^2}{2}.
\end{equation}
Hereafter equation (22) is referred to as the MSL relation.
Substituting the MSL relation into equation (13) and (15), we
obtain the following evolutionary equations:
\begin{equation}
\frac{dM}{dt}=f_{_{MSL}}(a_*,k,\lambda) {\dot M}_d,
\end {equation}
\begin{equation}
\frac{da_*}{dt}=M^{-1}g_{_{MSL}}(a_*,k,\lambda) {\dot M}_d,
\end{equation}
where $f_{_{MSL}}(a_*,k,\lambda)$ and $g_{_{MSL}}(a_*,k,\lambda)$
are CFBHM and CFBHS corresponding to the MSL relation,
respectively. In the same way, we obtain the parameter space
consisting of $a_*$ and $k$ as shown in Figure 15. The parameter
space is very dif\/ferent from that depicted in Figure 2, and it
is only divided into two regions by the boundary curve
$g_{_{MSL}}(a_*,k,\lambda)=0$, corresponding to only two states of
BH evolution. The characteristics of BH evolution related to the
MSL relation are shown in Tabel 3.

In this paper an analytic model of BH evolution is proposed for
GRBs. Although this model is rather particular and speculative ,
the potential application is still attractive: a little variation
of the angular velocity of the magnetic f\/ield lines might
af\/fect the output energy and evolutionary timescales of a BH
system remarkably. We also hope to improve and extend this model
to AGNs and stellar-mass BHs in  future work, since there is more
evidence to strongly suggest the existence of fast-spinning BHs
and possibly energy extraction via Blandford-Znajek-like
processes(Krolik 2001).

The anonymous referees are thanked for their suggestions on the
improvement of our manuscript. V. S. Beskin is thanked for his
encouragement and the discussion of the determination of
$\Omega_f$ in a BH magnetosphere. This work is supported by the
National Natural Science Foundation of China under grant No.
10173004.


\clearpage
\begin{deluxetable}{ccc}
\tablecolumns{3} \tablewidth{0pc} \tablecaption{BH EVOLUTIONARY
STATES CORRESPONDING TO FIVE DIFFERENT SUBREGIONS} \vspace{0.5cm}

\tablehead{
\colhead{}&\colhead{}&\colhead{}\\
\colhead{\begin{tabular}{cccc} $Region$&CFBHM&CFBHS&RP
Displacement
\end{tabular}}&\colhead{Evolutionary States}&\colhead{Terminal} \\
}

\startdata
&&\\
\begin{tabular}{cccc}
  \hspace{0.5cm} IA \hspace{1cm}&\hspace{0.5cm}$> 0$ \hspace{0.5cm}&\hspace{0.5cm} $< 0$ \hspace{0.5cm}&\hspace{0.2cm} towards the left \\ \\
  \tableline \\
   \hspace{0.5cm} IB \hspace{1cm}&\hspace{0.5cm}$> 0$ \hspace{0.5cm}&\hspace{0.5cm} $< 0$ \hspace{0.5cm}&\hspace{0.2cm} towards the left \\
\end{tabular} &
\begin{tabular}{c}increasing mass, \\ \\decreasing spin\\ \end{tabular}&
\begin{tabular}{c} equilibrium spin \\ \\ \tableline \\ zero spin \\ \end{tabular} \\ \\

\tableline \\
\begin{tabular}{cccc}
  \hspace{0.5cm} II \hspace{1cm}&\hspace{0.5cm}$> 0$ \hspace{0.5cm}&\hspace{0.5cm} $> 0$ \hspace{0.5cm}&\hspace{0.2cm} towards the right \\
\end{tabular} &
\begin{tabular}{c}
  increasing mass,\\increasing spin \\
\end{tabular} &
equilibrium spin \\ \\

\tableline \\
\begin{tabular}{cccc}
  \hspace{0.4cm} IIIA \hspace{0.9cm}&\hspace{0.5cm}$< 0$ \hspace{0.5cm}&\hspace{0.5cm} $< 0$ \hspace{0.5cm}&\hspace{0.2cm} towards the left \\ \\
  \tableline \\
   \hspace{0.4cm} IIIB \hspace{0.9cm}&\hspace{0.5cm}$< 0$ \hspace{0.5cm}&\hspace{0.5cm} $< 0$ \hspace{0.5cm}&\hspace{0.2cm} towards the left \\
\end{tabular} &
\begin{tabular}{c}decreasing mass, \\ \\decreasing spin\\ \end{tabular}&
\begin{tabular}{c} equilibrium spin \\ \\ \tableline \\ zero spin \\ \end{tabular} \\ \\

\enddata
\end{deluxetable}
\clearpage

\begin{deluxetable}{ccccccccc}

\tablecolumns{9} \tablewidth{0pc} \tablecaption{ $dE/dM_d(0)$
VARYING WITH $M_d(0)$ AND $a_*(0)$ FOR THE TWO SYSTEMS}
\tablehead{
\colhead{} &\colhead{} &\colhead{} &\colhead{} &\multicolumn{5}{c}{\begin{tabular}{c} \\ \hspace{3cm} $M_d(0)$ \hspace{3cm}  \\ \tableline \end{tabular}}\\
\colhead{$a_*(0)$} &\colhead{$k$} &\colhead{} &\colhead{} & \multicolumn{5}{c}{}\\
\colhead{} &\colhead{} & \colhead{} &\colhead{} & \colhead{0.05} &\colhead{0.10} &\colhead{0.15} &\colhead{0.20} &\colhead{0.25} }

\startdata
&&&&&&&&\\
0.2&0.5&&&0.1868&0.1849&0.1826&0.1796&0.1749  \\
&0.6&&&0.1827&0.1818&0.1808&0.1797&0.1784 \\
&&&&&&&&\\
0.3&0.5&&&0.1996&0.1989&0.1981&0.1974&0.1965\\
&0.6&&&0.19495&0.19489&0.19484&0.19478&0.19473\\
&&&&&&&&\\
0.4&0.5&&&0.2135&0.2133&0.2130&0.2127&0.2124\\
&0.6&&&0.2085&0.2088&0.2091&0.2094&0.2097\\
&&&&&&&&\\
0.5&0.5&&&0.2299&0.2298&0.2297&0.2296&0.2295\\
&0.6&&&0.2245&0.2250&0.2254&0.2259&0.2263\\
&&&&&&&&\\
0.6&0.5&&&0.2502&0.2501&0.2501&0.2500&0.2499\\
&0.6&&&0.2445&0.2450&0.2455&0.2460&0.2465\\
&&&&&&&&\\
0.8&0.5&&&0.3186&0.3175&0.3165&0.3155&0.3145\\
&0.6&&&0.31187&0.31193&0.31198&0.31203&0.31208\\
&&&&&&&&\\

\enddata
\end{deluxetable}

\clearpage
\begin{table}
\centerline{Table 3. BH EVOLUTIONARY STATES CORRESPONDING TO THE
MSL RELATION}
\begin{center}
\vspace{0.5cm}
\begin{tabular}{cccccc}
\tableline \tableline \\
Region & CFBHM & CFBHS & RP Displacement & Evolutionary States & Terminal \\ \\
\tableline \\
I & $> 0$ & $< 0$& towards the left & \begin{tabular}{c} increasing mass, \\ decreasing spin \end{tabular} & equilibrium spin \\ \\
\tableline \\
II & $> 0$ & $>0$& towards the right & \begin{tabular}{c} increasing mass, \\ increasing spin \end{tabular} & equilibrium spin \\ \\
\tableline
\end{tabular}
\end{center}
\end{table}

\clearpage

\begin{figure}
\plottwo{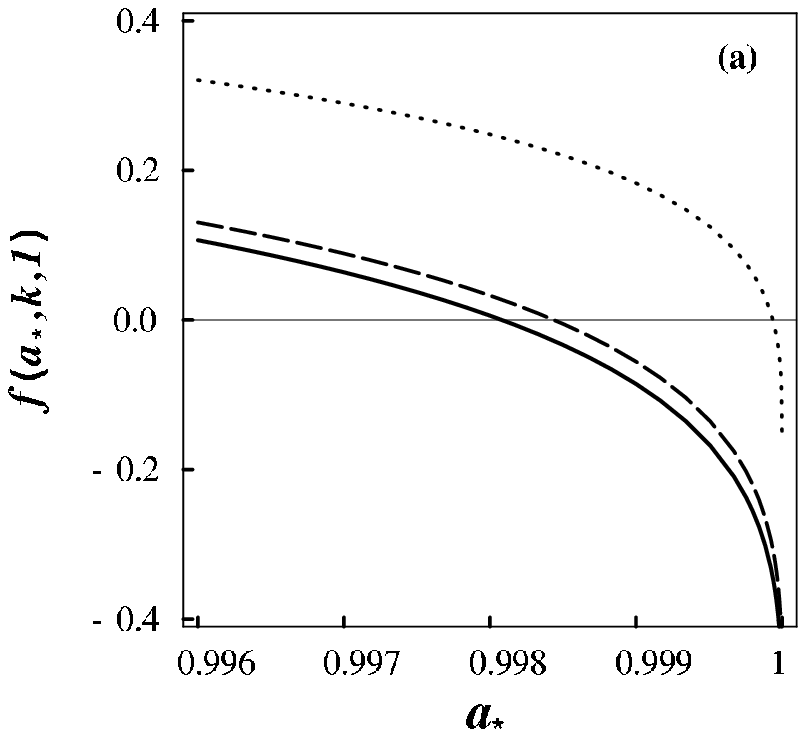}{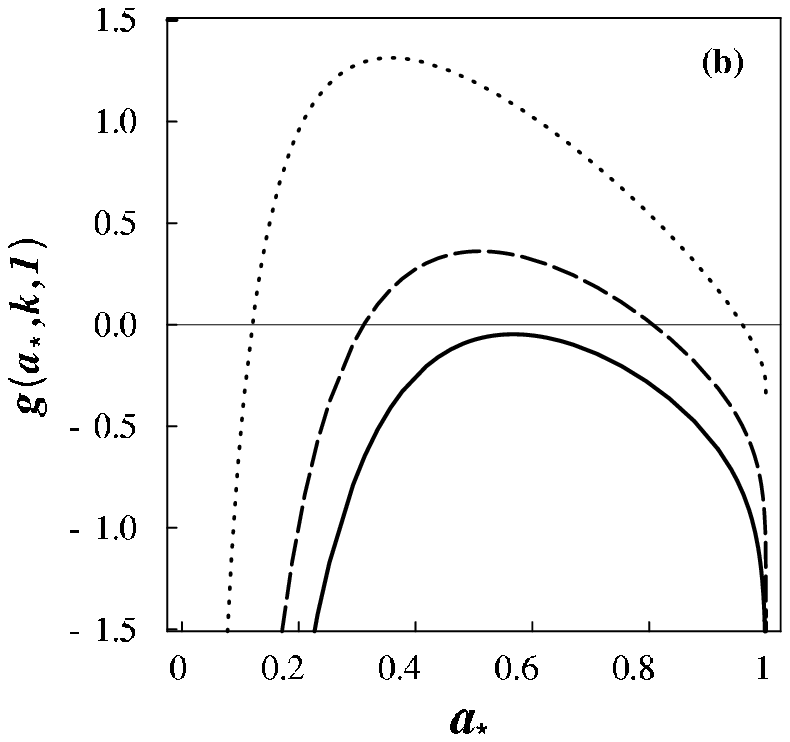}
\caption{(a) Curves of $f(a_*,k,\lambda)$ vs. $a_*$, which is
referred to as CFBHM to indicate the evolutionary characteristics
of the BH mass. (b) Curves of $g(a_*,k,\lambda)$ versus $a_*$,
which is referred to as CFBHS to indicate the evolutionary
characteristics of the BH spin. The parameters for the model are
$\lambda=1$, $k=0.5$ ({\it{solid line}}), $k=0.6$ ({\it{dashed
line}}) and $k=0.8$ ({\it{dotted line)}}.}
\end{figure}

\clearpage
\begin{figure}
\plottwo{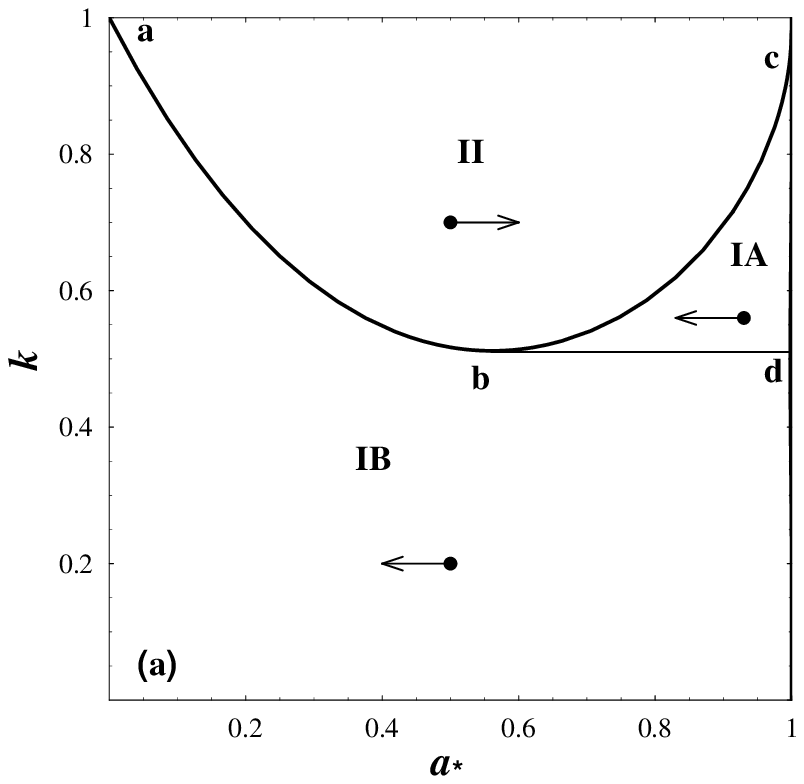}{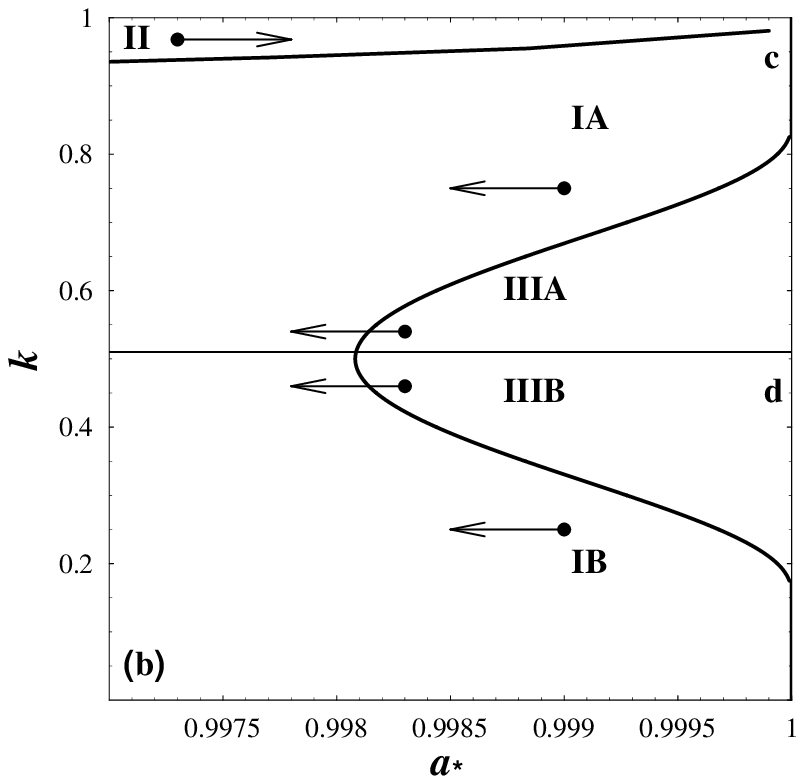}
\caption{Parameter space for BH evolution: the boundary curve of
regions I and II is represented by $g(a_*,k,\lambda)=0$, and the
boundary curve of regions I and III is represented by
$f(a_*,k,\lambda)=0$. BH evolutionary state is represented by the
filled circles with  arrows(RPs). It is noted  particularly that
the RP in region II moves towards the right, which means the BH
spin is increasing rather than decreasing in its evolutionary
process. The parameters for the model are $\lambda=1$ and $0<k<1$.
(a) $0<a_*<1$; (b) $ 0.997<a_*<1$ .}
\end{figure}

\clearpage
\begin{figure}
\epsscale{1} \plotone{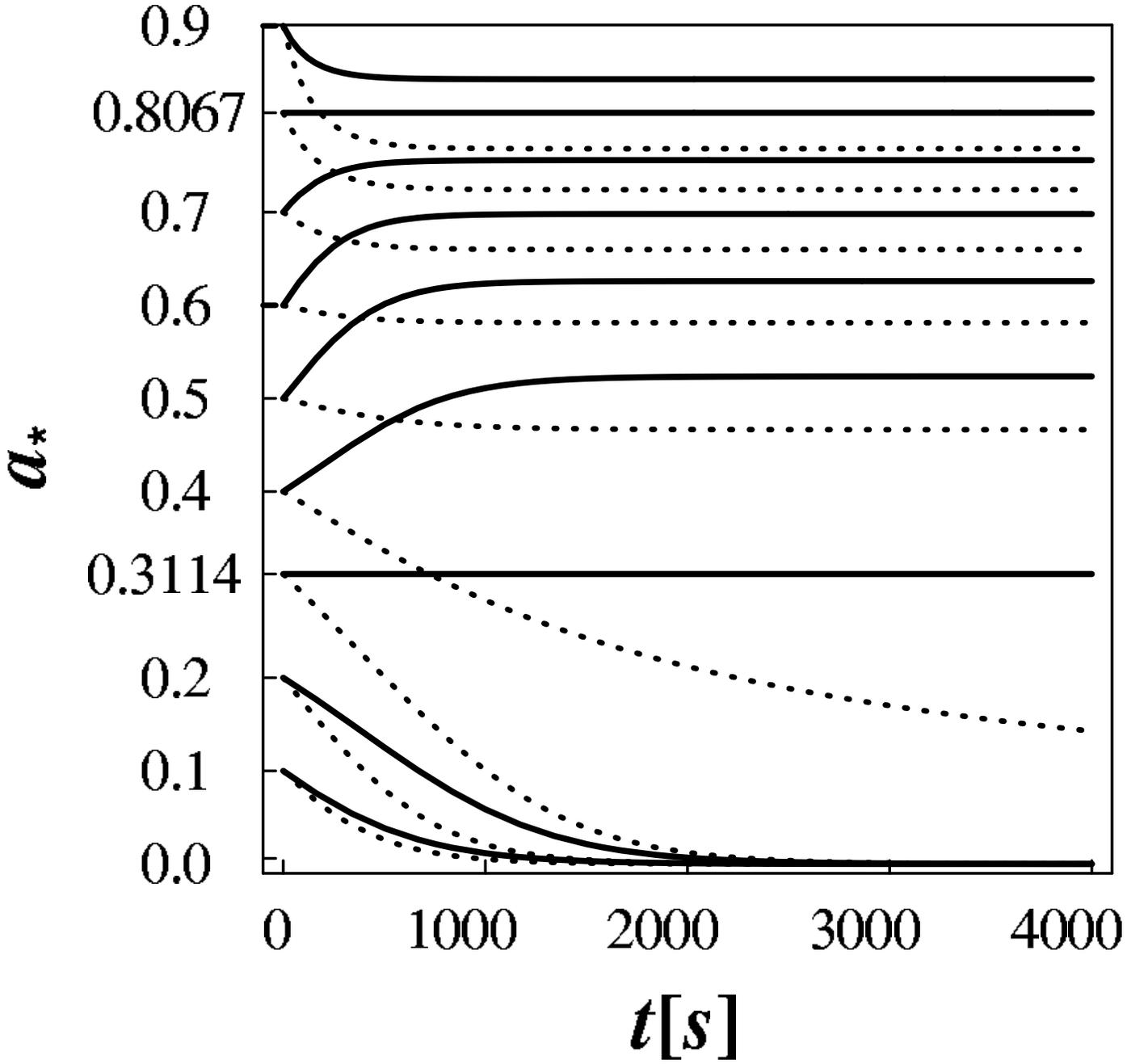} \caption{Evolution of the BH spin
with time in the units of seconds with solid lines for $k=0.6$ and
dashed lines for $k=0.5$. It is noted that the BH spin is
increasing with time as shown by the solid lines with
$0.3114<a_*(0)<0.8067$. The parameters for the model are
$M_h(0)=7M_\odot$, $M_d(0)=3M_\odot$ and $B_h(0)=10^{15}G$.}
\end{figure}

\clearpage
\begin{figure}
\epsscale{1} \plotone{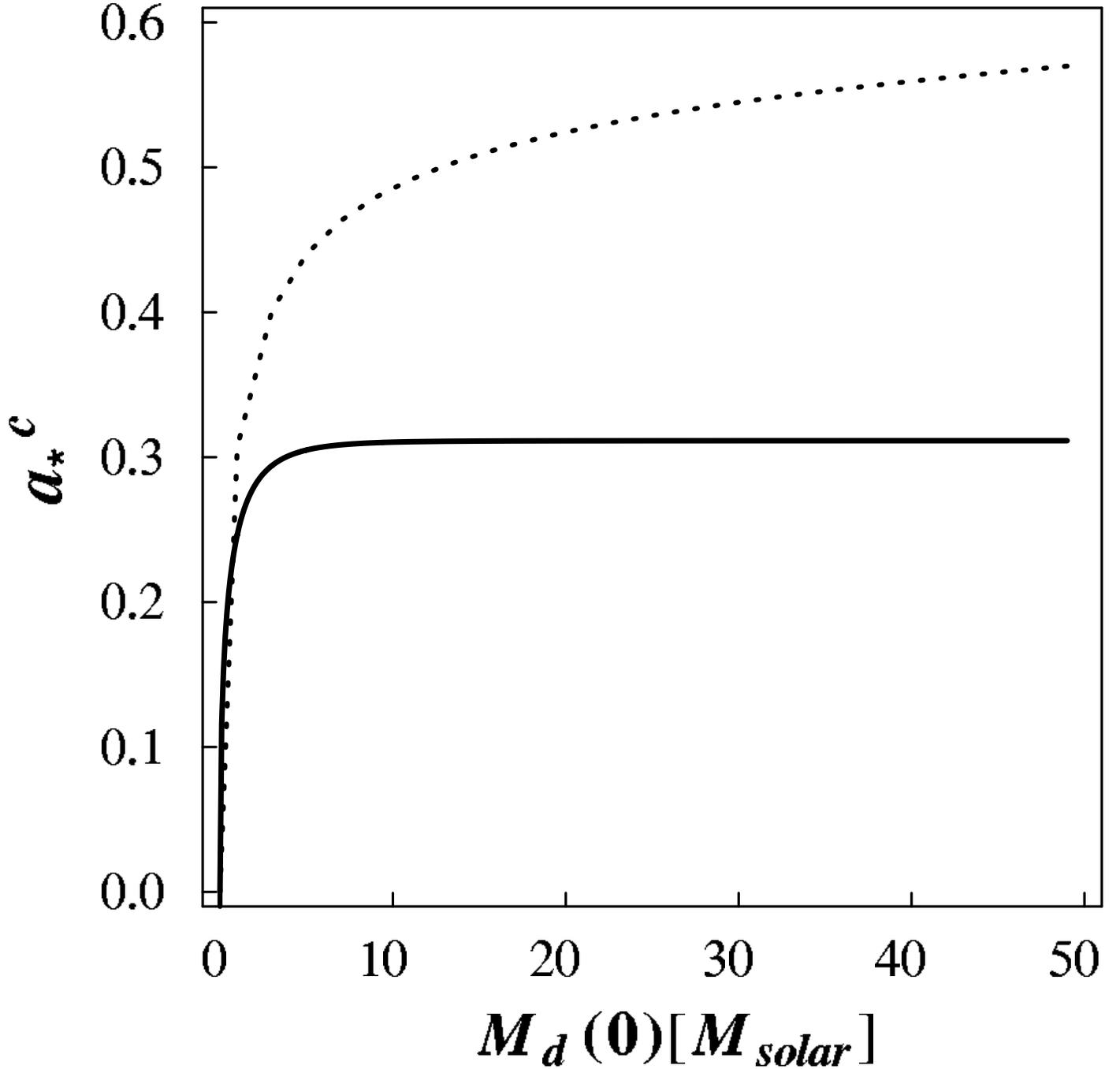} \caption{Critical BH spin $a_*^c$ to
ensure that the BH will stop spinning after all the initial disk
mass accretes onto the BH. It is noted that the solid line
($k=0.6$) is always below $a_*^c=0.3114$ for any values of
$M_d(0)$. The Parameters for the model are the same as for Fig.
3.}
\end{figure}

\clearpage
\begin{figure}
\epsscale{0.35}
\plotone{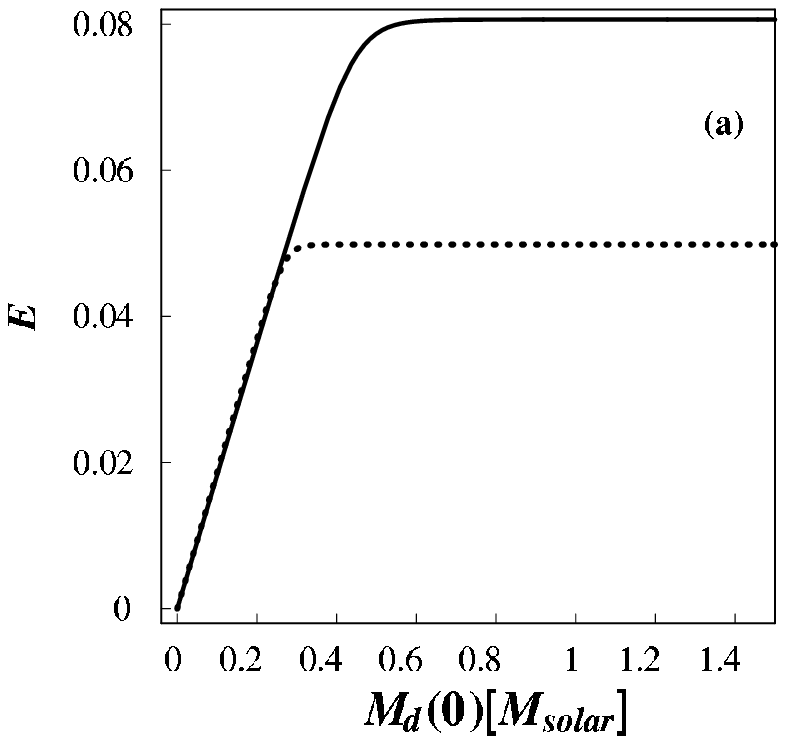}
\epsscale{0.34}
\plotone{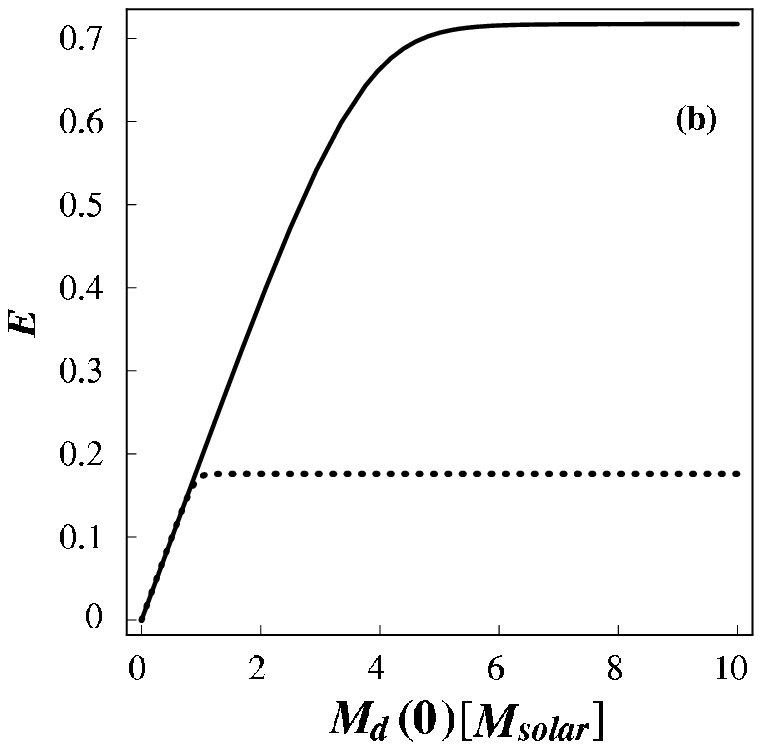}
\epsscale{0.34}
\plotone{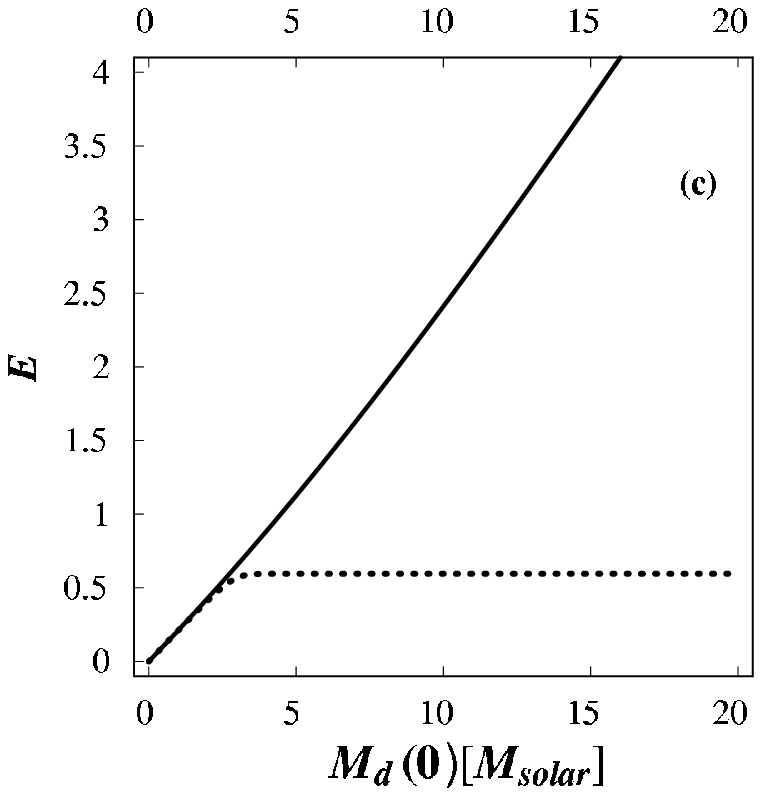}
\epsscale{0.33}
\plotone{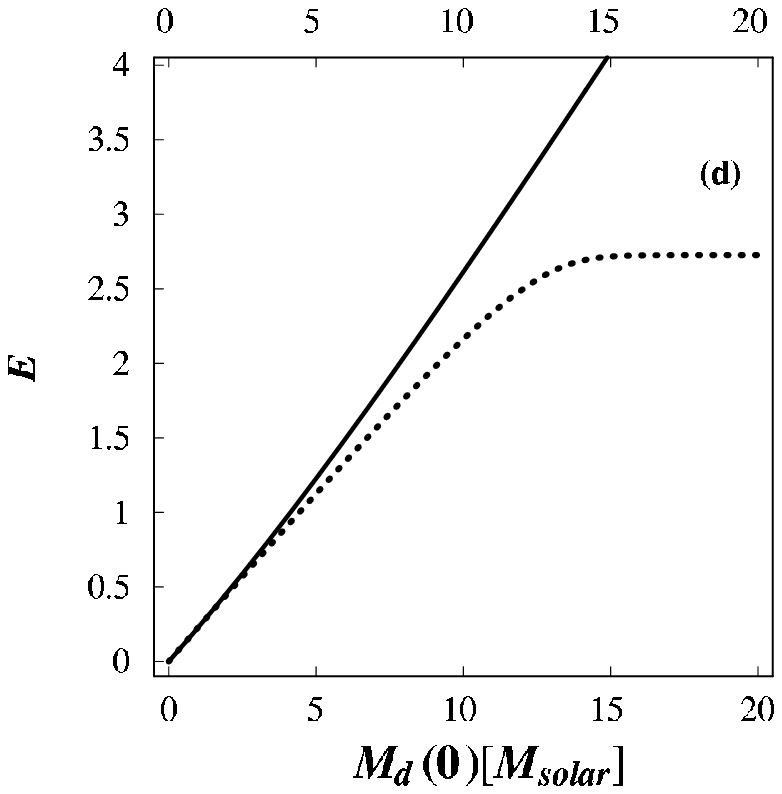}
\epsscale{0.35}
\plotone{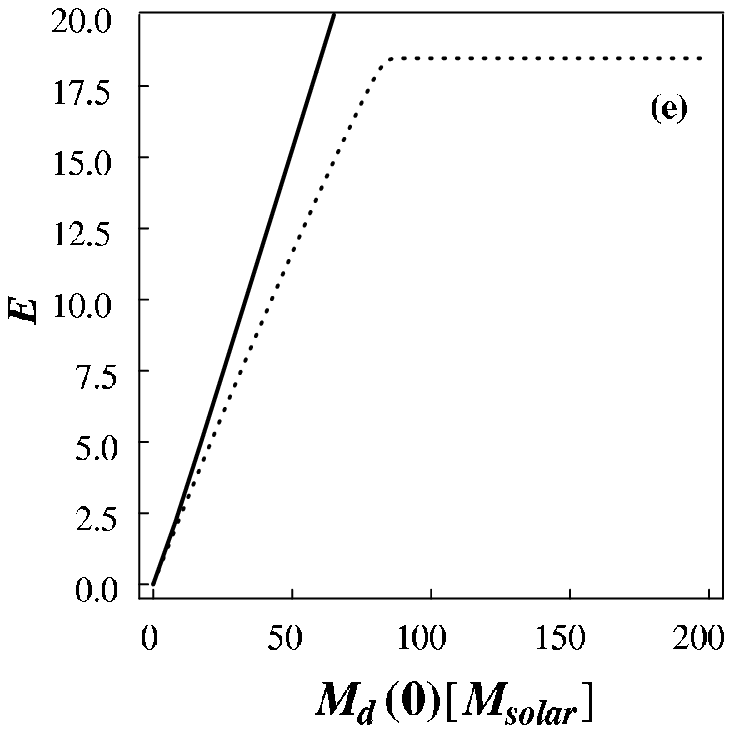}
\epsscale{0.34}
\plotone{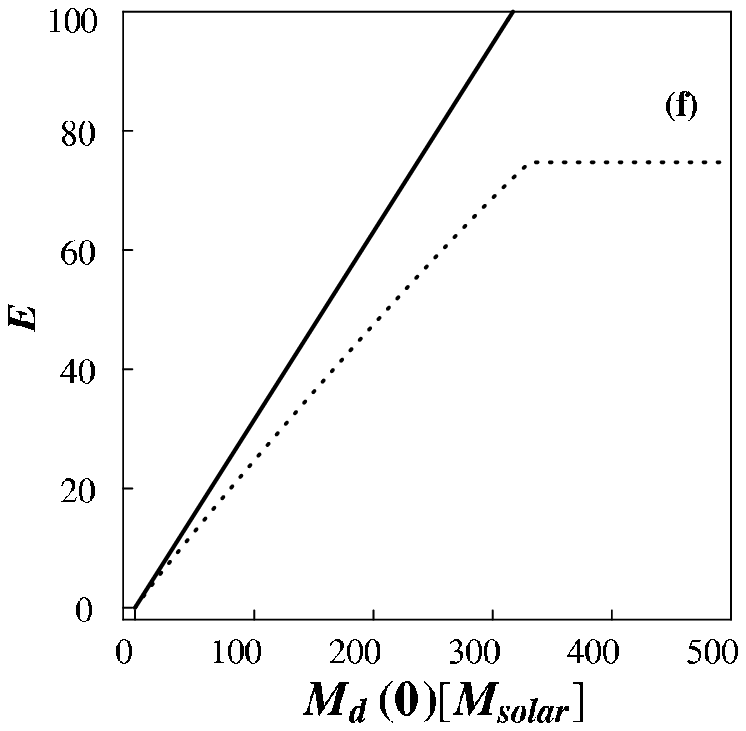}
\caption{Output energy vs. the initial disk mass with various
initial BH spin $a_*(0)$. It is noted that the solid line
($k=0.6$) is always higher than the dashed line($k=0.5$), and the
former has no upper limit for $a_*(0)>0.3114$. The parameters for
the model are  the same as for Fig.3 with $a_*(0)$ taken as (a)
0.2, (b) 0.3, (c) 0.4, (d) 0.5, (e) 0.6, (f) 0.8.}
\end{figure}

\clearpage
\begin{figure}
\epsscale{0.9} \plotone{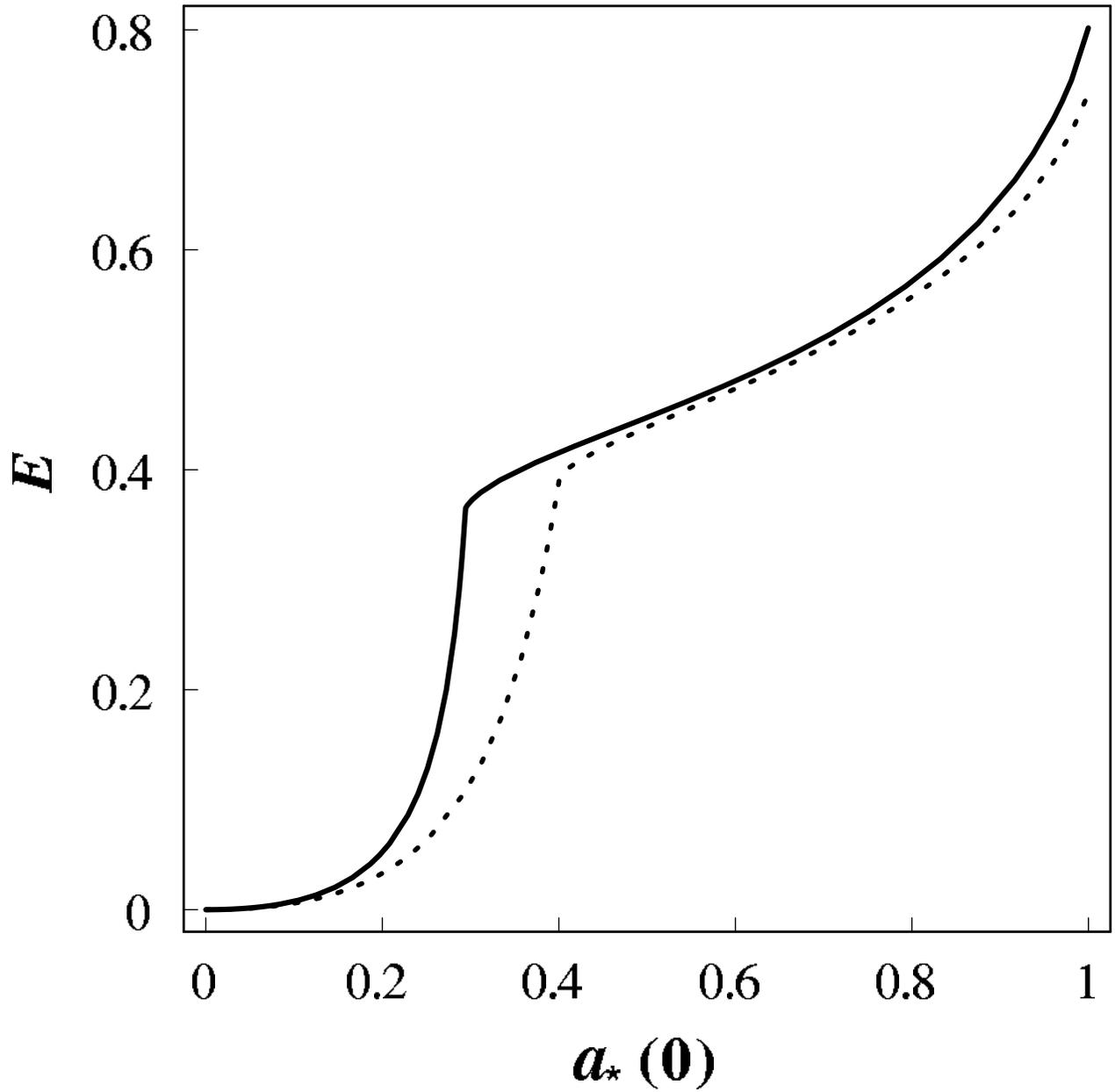} \caption{Upper limits of the
energy out of the system vs. the initial BH spin. It is noted that
the solid line ($k=0.6$) is always higher than the dashed
line($k=0.5$). The parameters for the model are the same as for
Fig.3.}
\end{figure}

\clearpage
\begin{figure}
\epsscale{0.9} \plotone{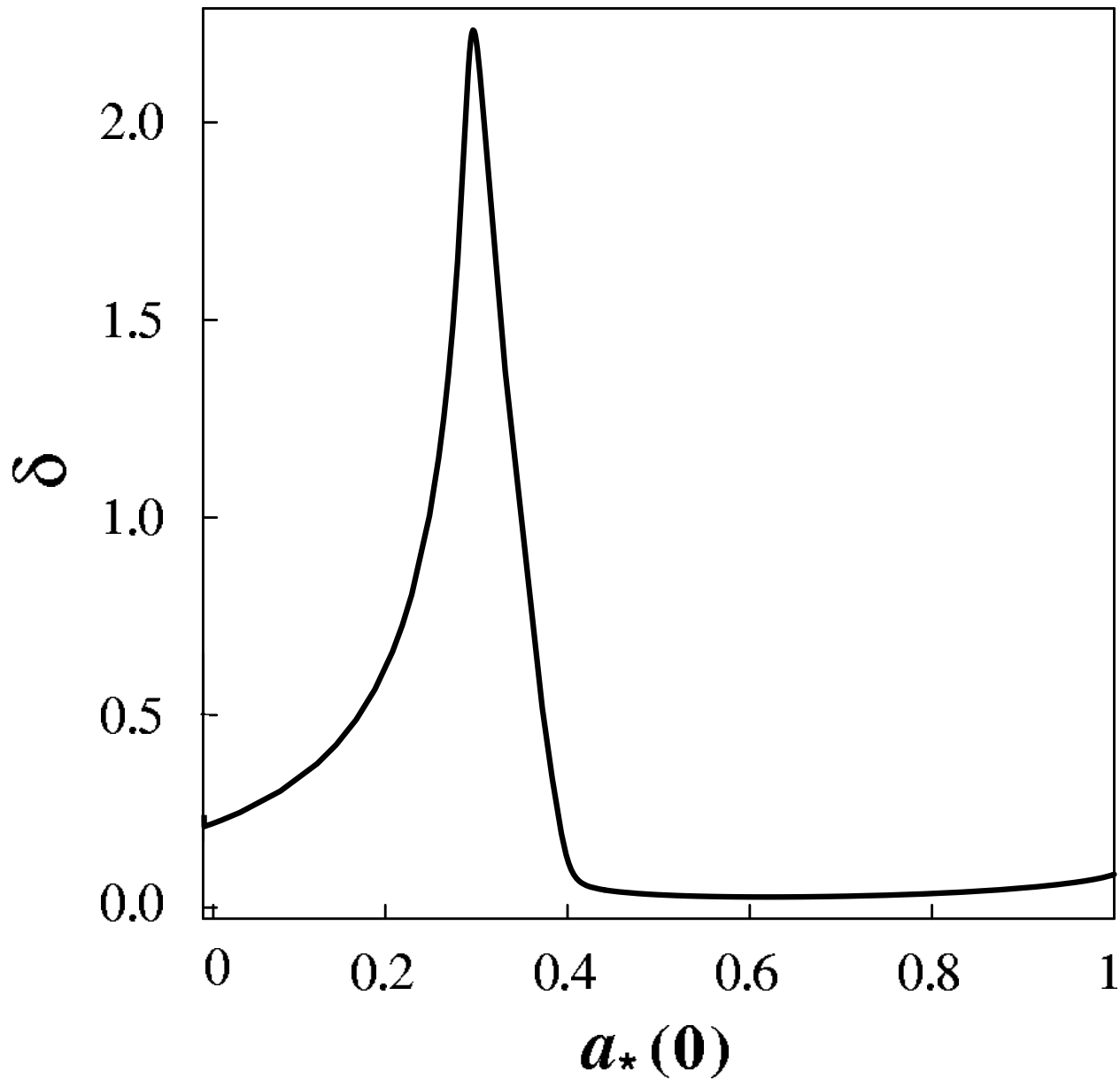} \caption{Relative dif\/ference of
output energy vs. the initial BH spin with its maximum
$\delta_{max} \approx 2.44$ at $a_*(0) \approx 0.294$. The
parameters for the model are the same as for Fig.3.}
\end{figure}

\clearpage
\begin{figure}
\epsscale{1} \plotone{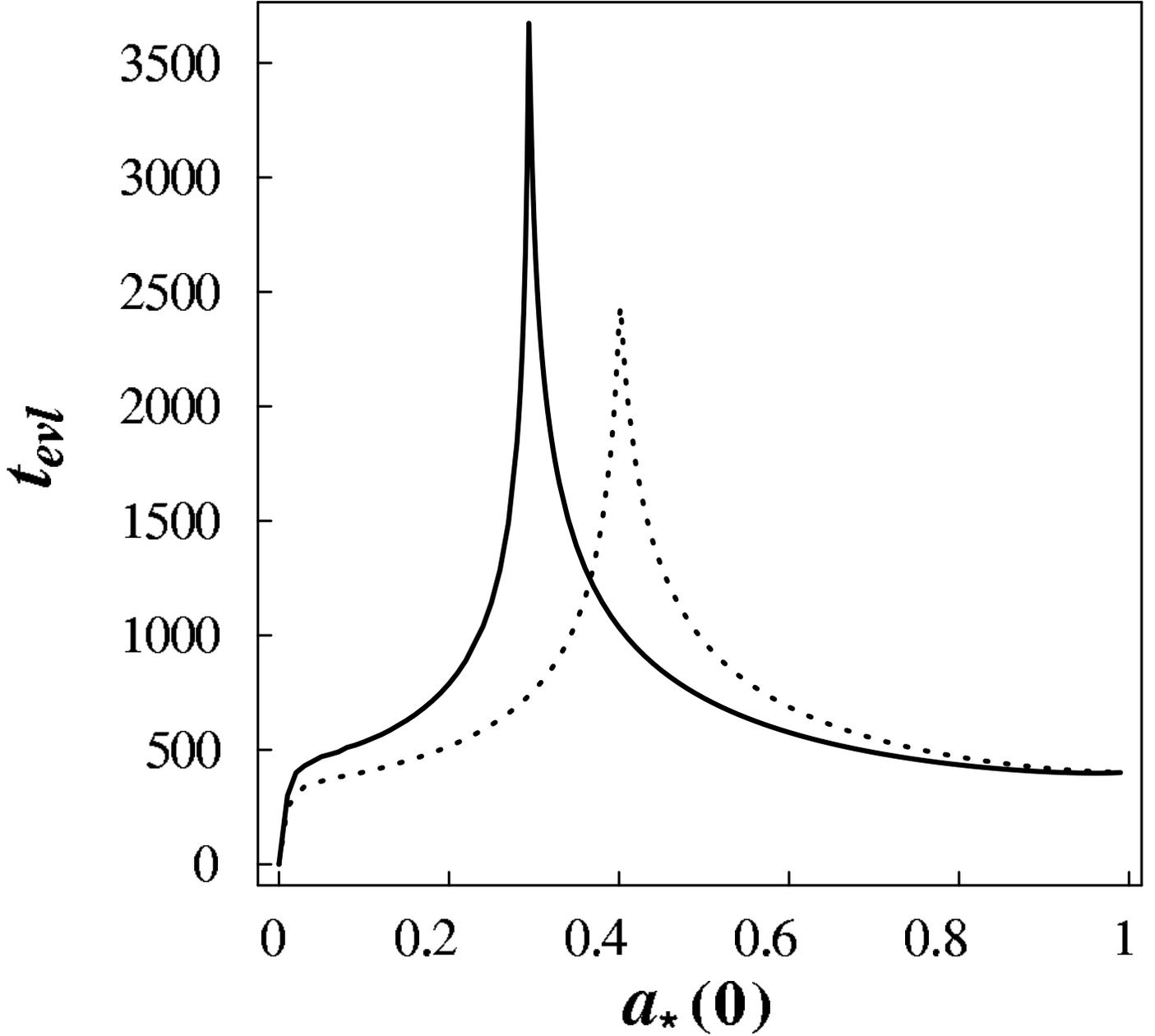} \caption{Evolutionary timescale of a
GRB powered by two BH systems with $k=0.6$ ({\it{solid line}}) and
$k=0.5$ ({\it{dashed line}}). It is noted that the former is less
than the latter for $a_*(0)>0.366$. The parameters for the model
are the same as for Fig.3 with the cutof\/f of $t_{evl}$ taken as
$T_{90}$ in the LK model.}
\end{figure}

\clearpage
\begin{figure}
\plottwo{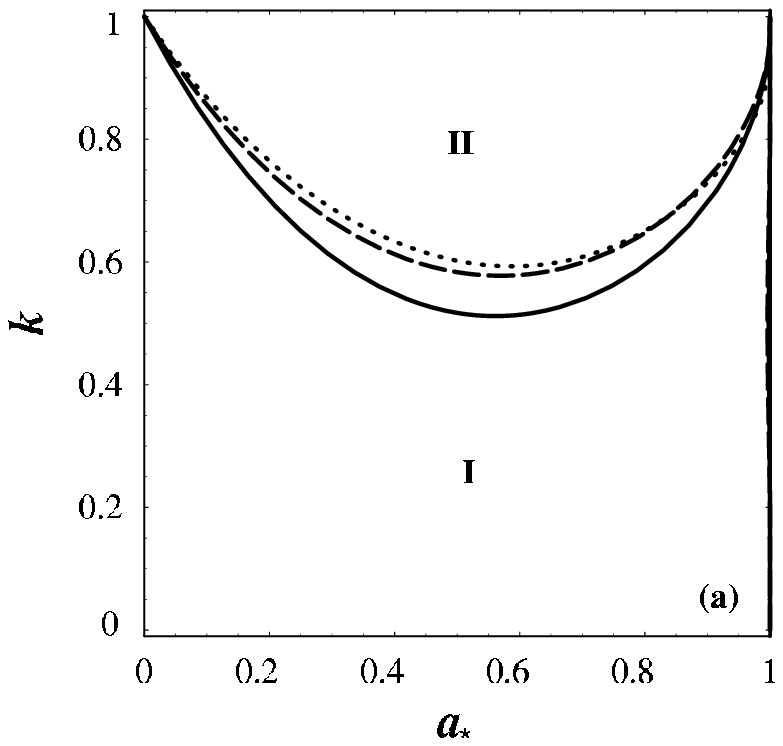}{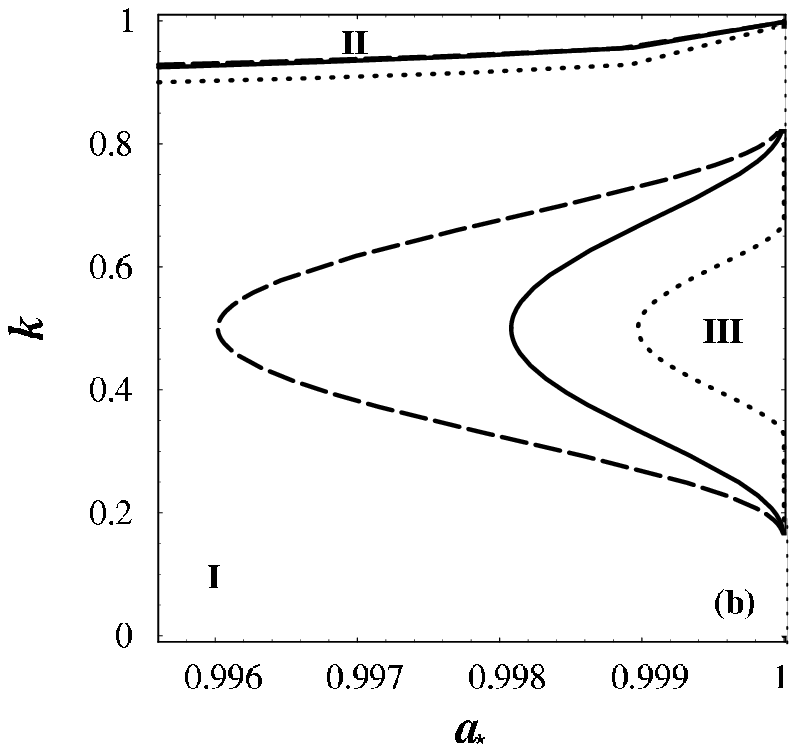}
\caption{Parameter space for BH evolution corresponding to
dif\/ferent values of $ \lambda$. It is noted that the boundary
curves vary nonmonotonically with $\lambda$ . The parameters for
the model are $k=0.6$, $\lambda=1$ ({\it{solid line}}),
$\lambda=0.5$ ({\it{dashed line}}), and $\lambda=0$ ({\it{dotted
line}}) with (a) $0<a_*<1$, (b) $0.9956<a_*<1$. }
\end{figure}

\clearpage
\begin{figure}
\epsscale{0.9} \plotone{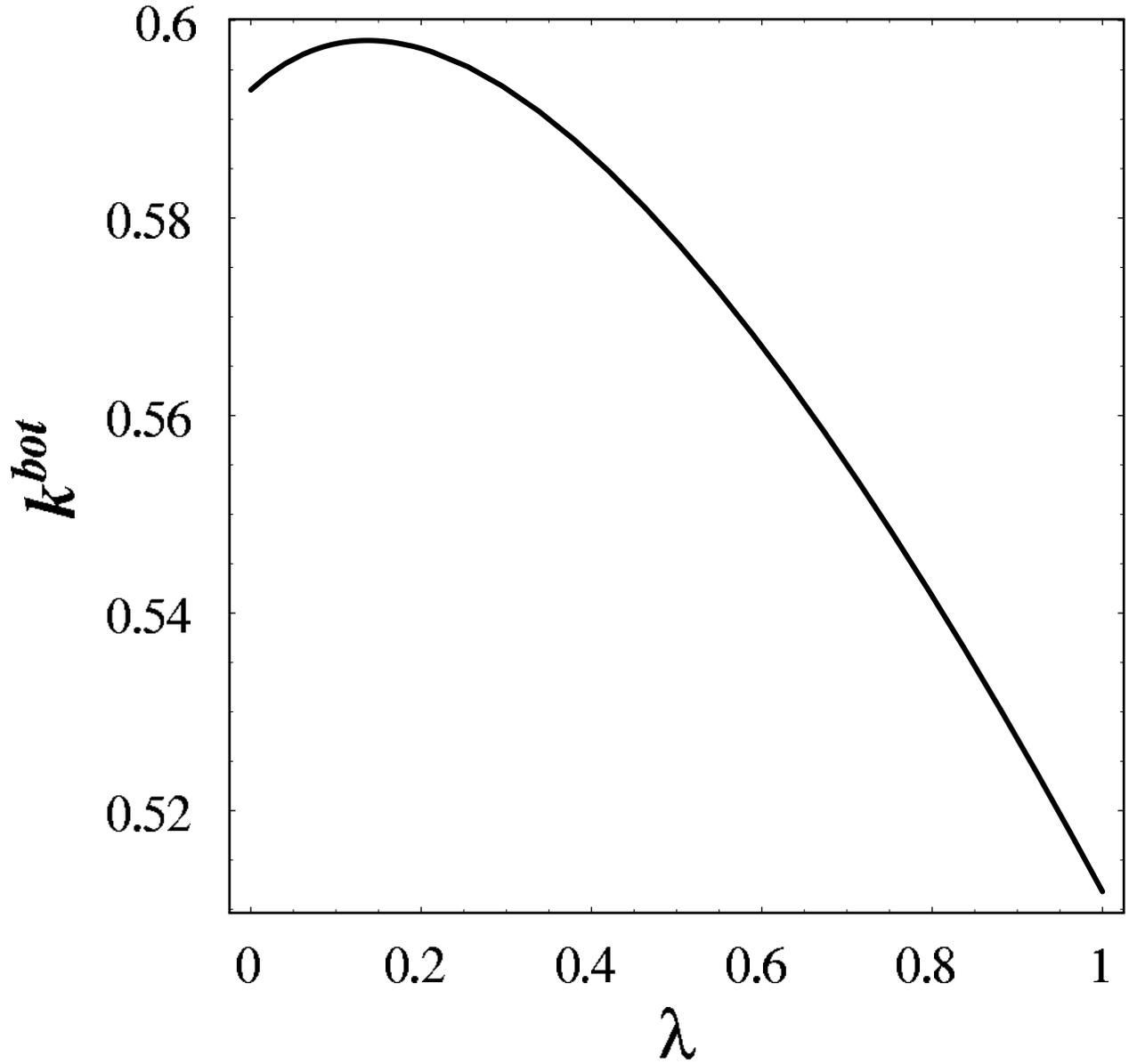} \caption{Vertical ordinate
$k^{top}$ of the bottom point of CFBHS varying with $\lambda$ for
$0 \le \lambda<1$.}
\end{figure}

\clearpage
\begin{figure}
\epsscale{0.9} \plotone{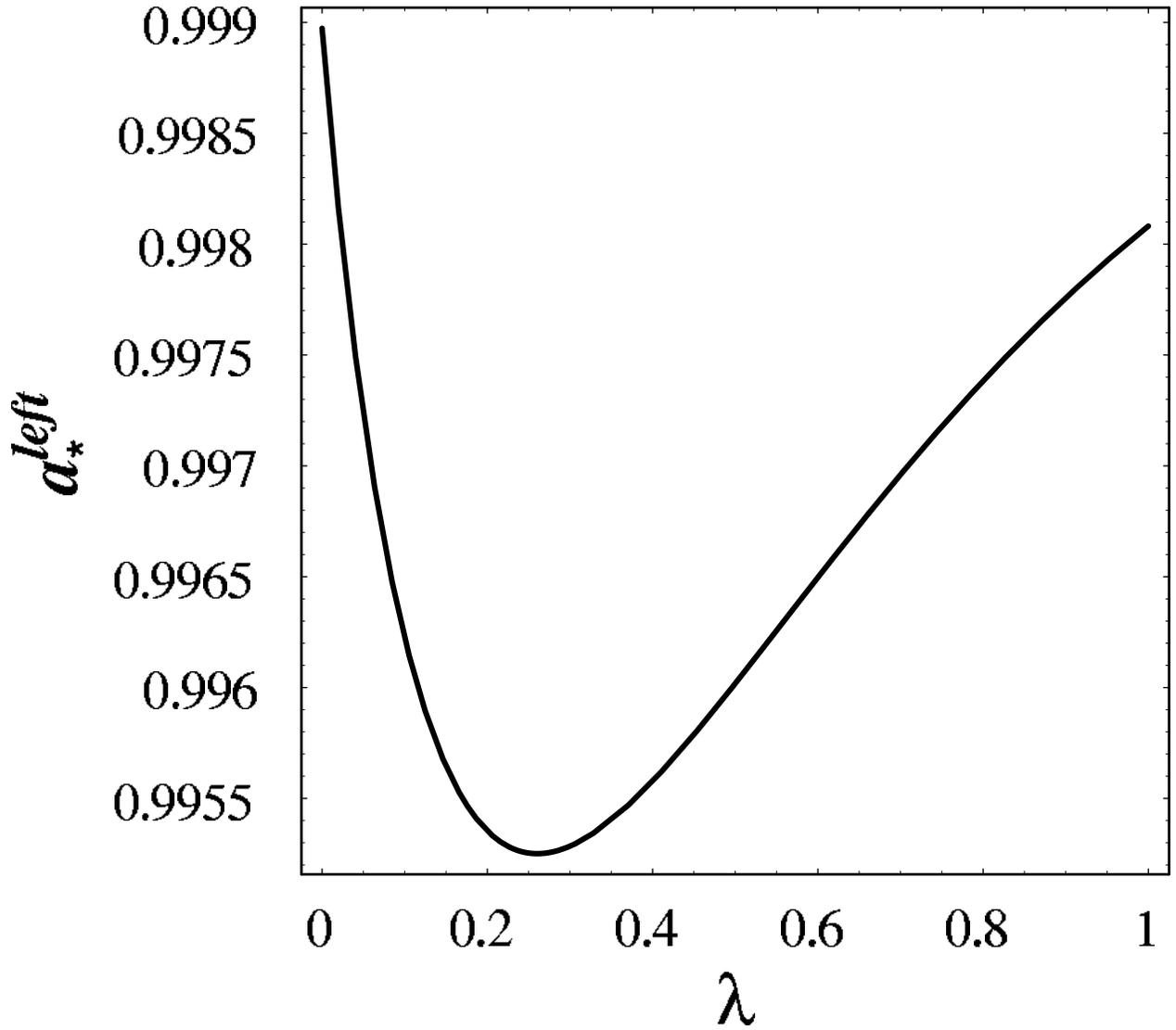} \caption{Horizontal ordinate
$a_*^{left}$ of the leftmost point of CFBHM varying with $\lambda$
for $0 \le \lambda<1$.}
\end{figure}

\clearpage
\begin{figure}
\epsscale{1.1} \plotone{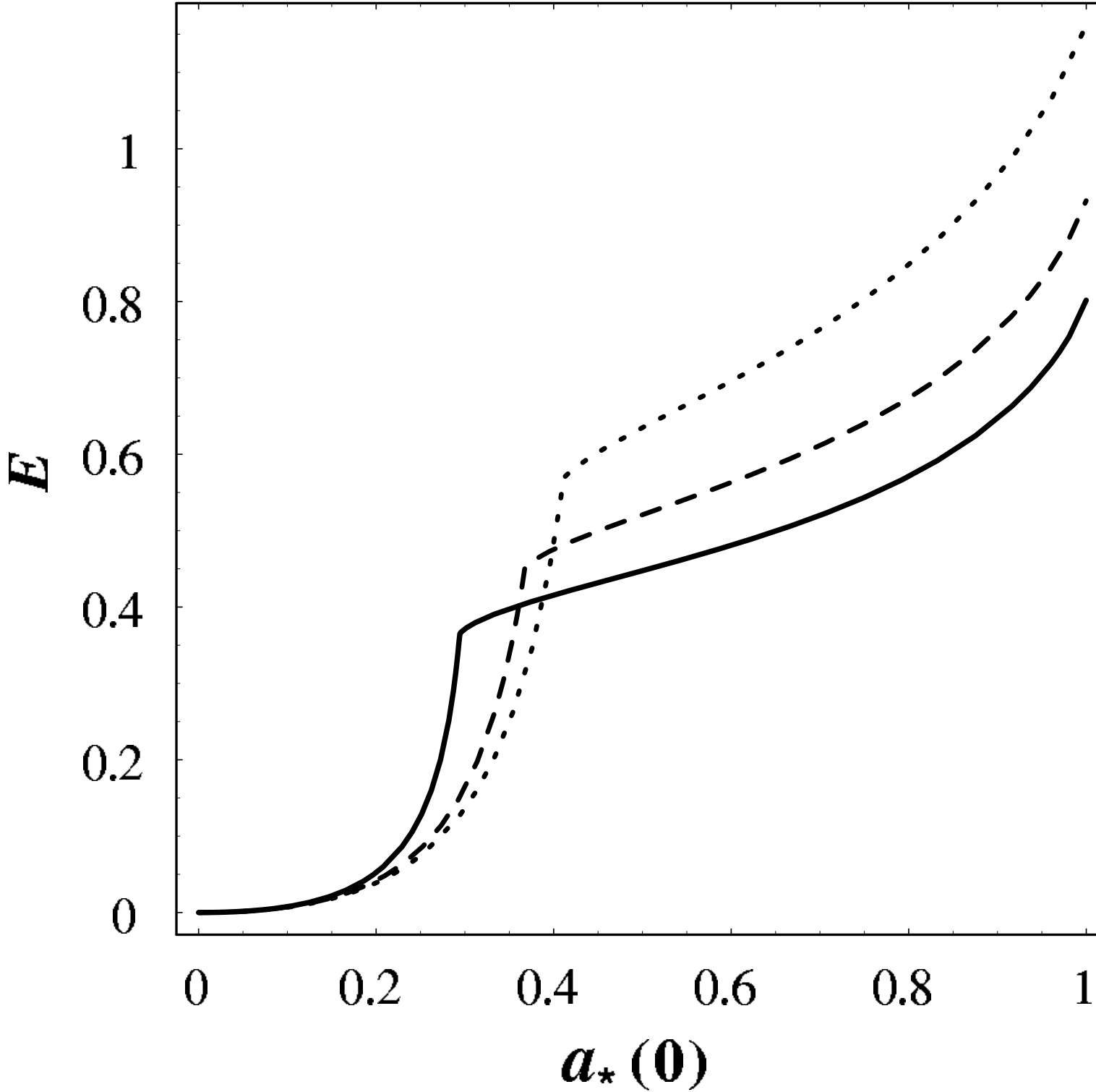} \caption{Upper limits of the
energy out of the system vs. the initial BH spin. It is noted that
the output energy from a system of a slow-spinning BH with a thin
disk might be greater than that from a system with a thick disk,
while the output energy from a system of a fast-spinning BH with a
thin disk might be less than that from a system with a thick disk.
The parameters for the model are $k=0.6$, $\lambda=1$ ({\it{solid
line}}), $\lambda=0.5$ ({\it{dashed line}}), and $\lambda=0$
({\it{dotted line}}).}
\end{figure}

\clearpage
\begin{figure}
\epsscale{1} \plotone{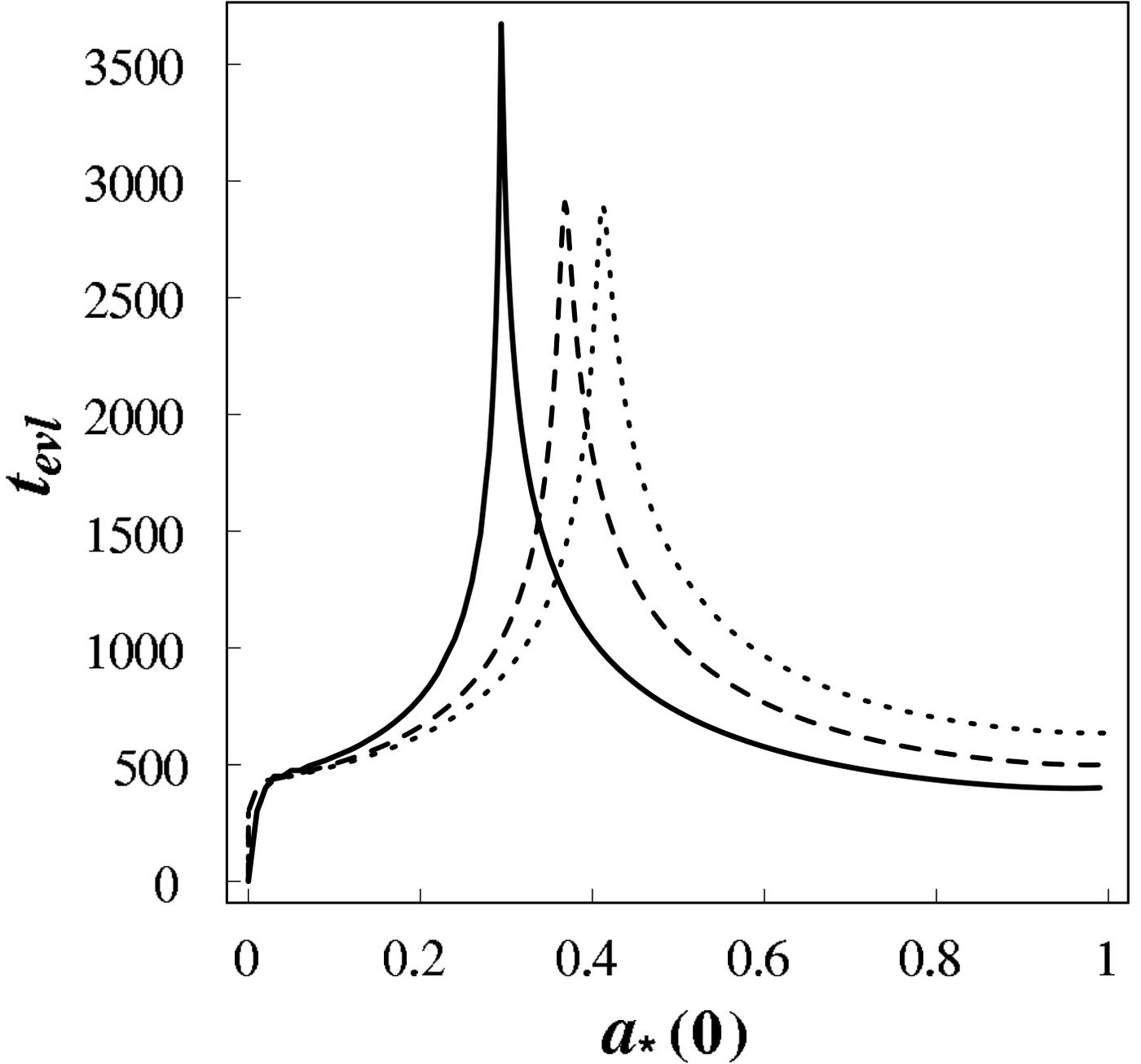} \caption{Evolution time of a GRB
vs. the initial BH spin. It is noted that the evolutionary
timescale of a slow-spinning BH with a thin disk might be longer
than that with a thick disk, while the evolutionary timescale of a
fast-spinning BH with a thin disk might be less than that with a
thick disk. The parameters for the model are the same as for
Fig.12.}
\end{figure}

\clearpage
\begin{figure}
\epsscale{1} \plotone{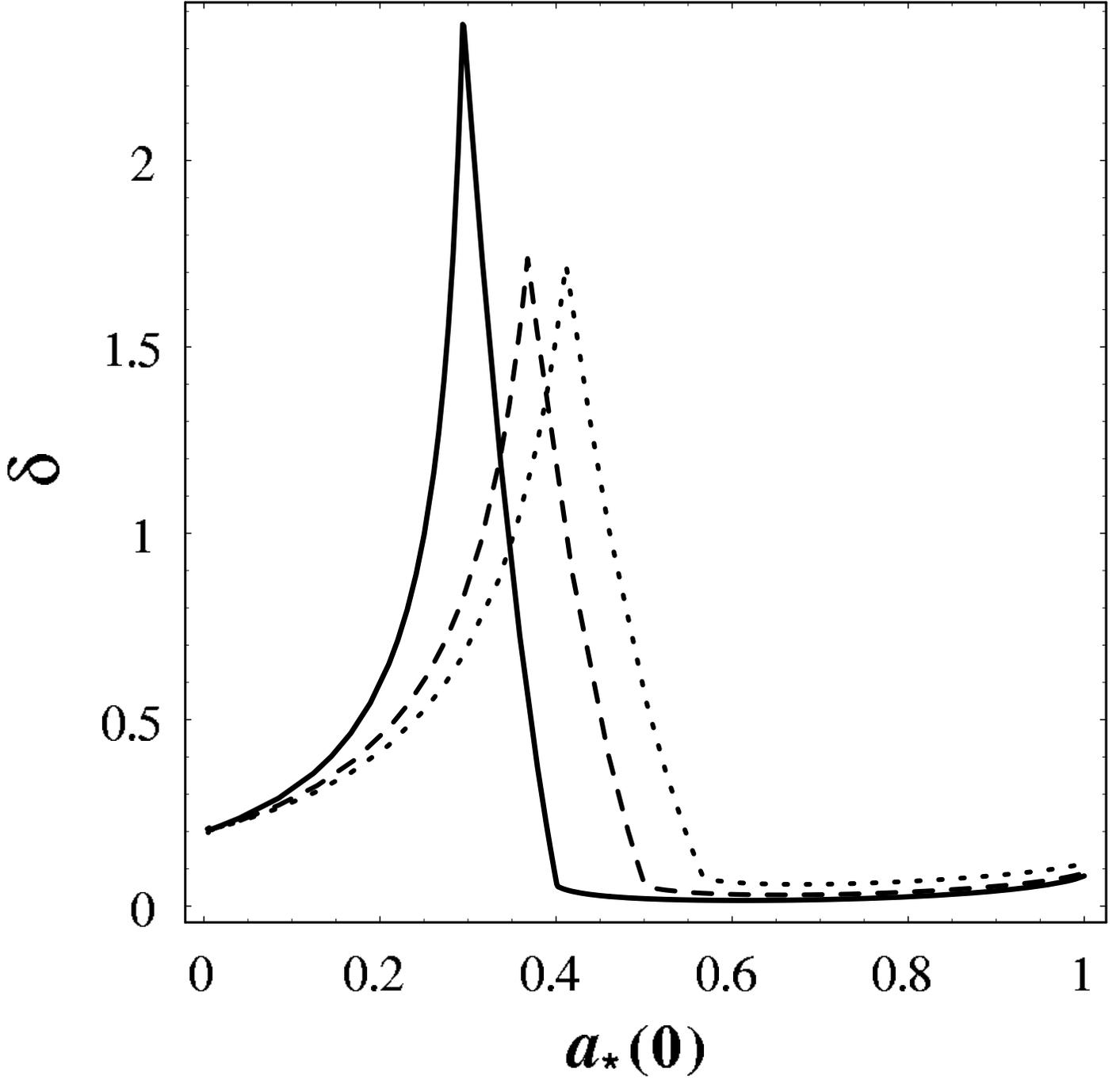} \caption{Relative dif\/ference
between the output energy from the system with $k=0.6$ and that
with $k=0.5$ for dif\/ferent values of $\lambda$. It is noted that
$\delta$ varies nonmonotonically with $a_*(0)$  for dif\/ferent
peak values corresponding to  dif\/ferent values of $\lambda$ and
$a_*(0)$. The parameter for the model are the same as for Fig.12.}
\end{figure}

\clearpage
\begin{figure}
\epsscale{1} \plotone{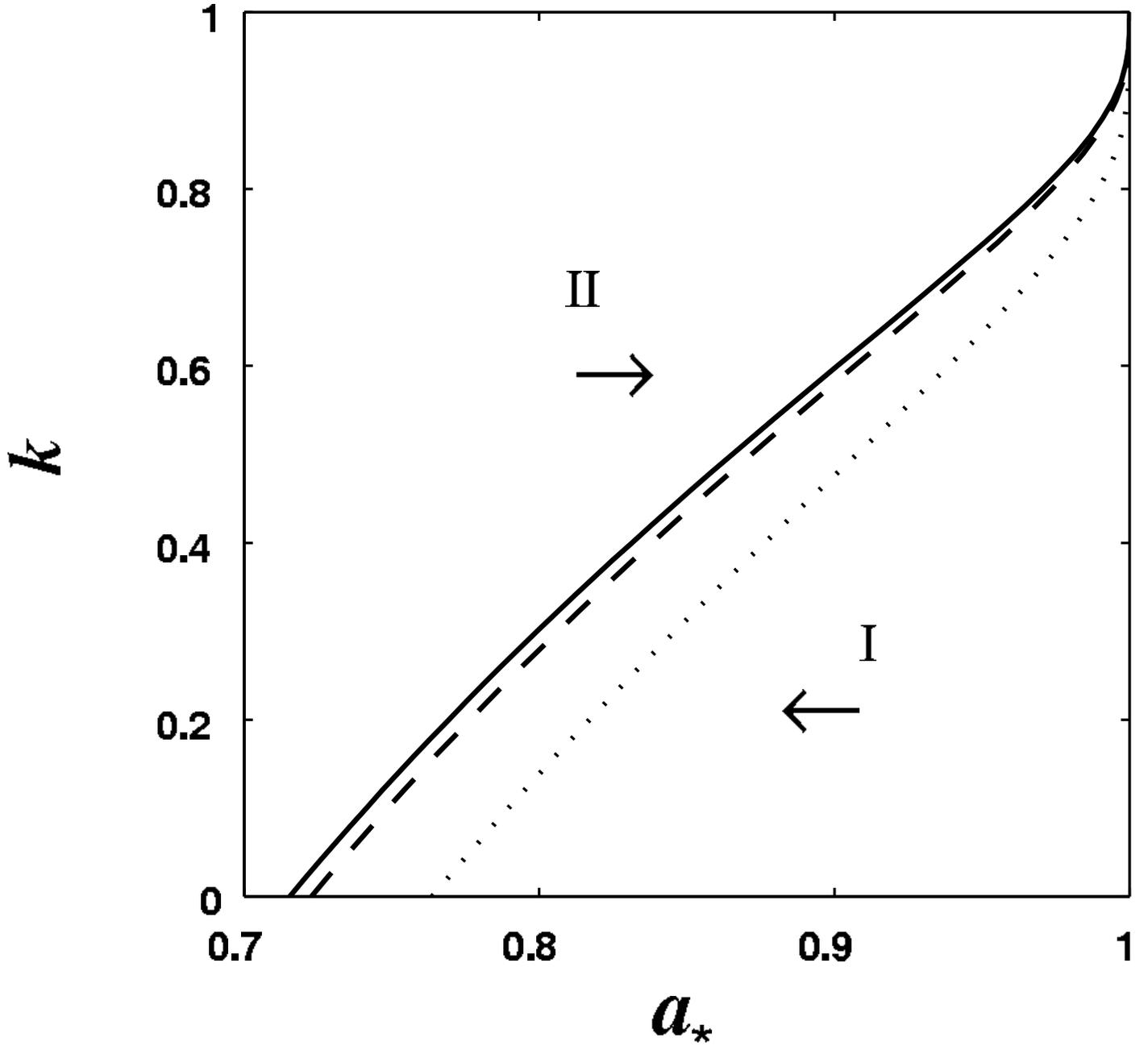} \caption{Parameter space for BH
evolution corresponding to the MSL relation with  dif\/ferent
values of $\lambda$. The parameters for the model are $ \lambda=1$
({\it{solid line}}), $ \lambda=0.5$ ({\it{dashed line}}),
$\lambda=0$ ({\it{dotted line}}).}
\end{figure}

\end{document}